\documentclass[a4paper,11pt]{article}
\usepackage{amsmath}
\usepackage{amssymb}
\usepackage{amsfonts}
\usepackage{graphicx}
\usepackage{graphics}
\usepackage{float}
\usepackage{dsfont}
\usepackage{a4wide}
\usepackage{subfig}
\usepackage{feynmp}
\usepackage{indentfirst}

\newcommand{\Sp}{\mathrm{Sp}}
\newcommand{\bs}[2]{{{\sf{b}}
\phantom{]}}^{\!\!\!\!\!\!*{\kern 1.3pt}{\mbox{${\scriptstyle #1}$}}}_{\mbox{${\scriptstyle #2}$}}}
\newcommand{\as}[2]{{{\sf{a}}
\phantom{]}}^{\!\!\!\!\!\!*{\kern 1.3pt}{\mbox{${\scriptstyle #1}$}}}_{\mbox{${\scriptstyle #2}$}}}
\newcommand{\naf}[2]{n'^{\!\!\!\!\!*{\kern 2.8pt}{\mbox{${\scriptstyle(#1)}$}}}
_{\mbox{${\scriptstyle #2}$}}}
\newcommand{\na}[2]{n^{\!\!\!\!*{\kern 1.3pt}{\mbox{${\scriptstyle(#1)}$}}}
_{\mbox{${\scriptstyle #2}$}}}
\newcommand{\ap}[2]{a^{\!\!\!\!*{\kern 1.3pt}{\mbox{${\scriptstyle(#1)}$}}}
_{\mbox{${\scriptstyle #2}$}}}
\newcommand{\ep}[1]{e^{\!\!\!\!*{\kern 1.3pt}{\mbox{${\phantom{()}}$}}}
_{\mbox{${\scriptstyle #1}$}}}
\newcommand{\ad}[2]{a^{\!\!\!\!*{\kern 1.3pt}{\mbox{${\scriptstyle #1}$}}}_{\mbox{${\scriptstyle #2}$}}}

\newcommand {\bb}[1]{\mbox{\boldmath $#1$}}
\newcommand{\ii}{\mathrm{i}}

\begin{document}

\title{\bf Neutrino oscillations in
dense matter}
\author{\bf A.~E.~Lobanov\thanks{E-mail: lobanov@phys.msu.ru}}

\date{}

\maketitle
\begin{center}
{\em {Department of Theoretical Physics, Faculty of Physics,
  Moscow State University, 119991 Moscow, Russia }}
\end{center}

\begin{abstract}
We propose a modification of the electroweak
theory, where the fermions with the same
electroweak quantum numbers are  combined in   multiplets and
are treated as different quantum states of a single particle.
The developed  approach enables one  to calculate the
probabilities of the processes taking place in the  detector at
long distances from the  particle source.
Calculations of higher-order processes including the computation of the contributions due to radiative corrections can be performed in the framework of
the perturbation theory using the regular diagram technique. As a result, we obtain
the analog to the Dirac--Schwinger equation of quantum electrodynamics
describing neutrino oscillations and its spin rotation in  dense matter.
\end{abstract}

In paper \cite{Lobanov_2} the Hilbert
spaces of particle states are constructed in such a way that
the neutrinos, the charged leptons and the down- and
up-type quarks  are combined in $SU(3)$-multiplets with their components
being considered as different quantum states of a single particle.

The Lagrangian for the physical fermion fields in the Standard Model modified in accordance with these considerations is written as
\begin{equation}\label{l2}
{\cal{L}}_{f}={\cal{L}}_{0}+{\cal{L}}_{{\mathrm{int}}},
\end{equation}
\noindent where
\begin{equation}\label{l3}
\displaystyle {\cal{L}}_{0}= \frac{\ii}{2}\sum\limits_{i=\nu, l,u,d}\left[\left(\bar{\varPsi}^{(i)}
\gamma^{\mu}
(\partial_{\mu}{\varPsi}^{(i)})\right)
-(\partial_{\mu}\bar{\varPsi}^{(i)})
\gamma^{\mu}{\varPsi}^{(i)}
   \right]\;-\!\!\!\!\!\sum\limits_{i=\nu, l,u,d}\!\!\!
   \bar{\varPsi}^{(i)}{\mathds{M}}^{(i)}{\varPsi}^{(i)}
\end{equation}
\noindent is the Lagrangian of free fields and
\begin{equation}\label{l1}
\begin{array}{c}
\displaystyle {\cal{L}}_{{\mathrm{int}}}= -\!\!\!\!\!\sum\limits_{i=\nu, l,u,d}\!\!\!\bar{\varPsi}^{(i)}
 {\mathds{M}}^{(i)}({H}/{v}){\varPsi}^{(i)}\\
\displaystyle -\frac{g}{2\sqrt{2}}
\left(\bar{\varPsi}^{(l)}\gamma^{\mu}
(1+\gamma^{5})\,{\varPsi}^{(\nu)}
W_{\mu}^{-}+\bar{\varPsi}^{(\nu)}\gamma^{\mu}
(1+\gamma^{5})\,{\varPsi}^{(l)}
W_{\mu}^{+}\right)\\[12pt]
\displaystyle -\frac{g}{2\sqrt{2}}
\left(\bar{\varPsi}^{(d)}\gamma^{\mu}
(1+\gamma^{5})\,{\varPsi}^{(u)}
W_{\mu}^{-}+\bar{\varPsi}^{(u)}\gamma^{\mu}
(1+\gamma^{5})\,{\varPsi}^{(d)}
W_{\mu}^{+}\right)\\
\displaystyle -e\!\!\!\sum\limits_{i= l,u,d}Q^{(i)}\bar{\varPsi}^{(i)}\gamma^{\mu}\,
  {\varPsi}^{(i)}A_{\mu}\\
\displaystyle -\frac{g}{2\cos\theta_{\mathrm{W}}}\!\sum\limits_{i=\nu, l,u,d}\!\!
\bar{\varPsi}^{(i)}\gamma^{\mu}
\left(T^{(i)}-2Q^{(i)}\sin^{2}
\theta_{\mathrm{W}}
+T^{(i)}\gamma^{5}\right)
{\varPsi}^{(i)}Z_{\mu}
\end{array}
\end{equation}
\noindent is the interaction Lagrangian  between the
fermion fields,  the vector boson fields $W^{\pm}_{\mu},
Z_{\mu}, A_{\mu}$, and the Higgs field $H$. Here
$\theta_{\mathrm{W}}$ is the Weinberg angle,
$e=g\sin\theta_{\mathrm{W}}$ is the positron electric charge,
$T^{(i)}$ is the weak isospin ($T^{(\nu)}=T^{(u)}= 1/2,\,
T^{(l)}=T^{(d)}= -1/2$), $Q^{(i)}$ is the electric charge of the
multiplet in the units of $e$. The value $v$ is the vacuum
expectation of the Higgs field.

Thus, this Lagrangian  formally coincides with the Lagrangian of the
Standard Model supplemented by  right-handed neutrino singlet (see, e.g., \cite{pdg}).
However,  the wave
functions $ {\varPsi}^{(i)}$ describe not the individual
particles, but the multiplets as a whole{\footnote{For neutrinos $(i)=(\nu)$, for charged leptons $(i)=(l)$, for up-type quarks $(i)=(u)$, for down-type quarks  $(i)=(d).$}}. So it is not
necessary to introduce the mixing matrices into
${\cal{L}}_{{\mathrm{int}}}$ explicitly.
The field functions for the fermion fields represent 12-component objects which satisfy the modified Dirac equations
\begin{equation}\label{6}
\left( \ii\gamma^\mu\partial_\mu{\mathds{I}}  -{\mathds{M}}^{(i)}\right){\varPsi}^{(i)}(x)=0.
\end{equation}
\noindent  In these equations ${\mathds{M}}^{(i)}$ are the Hermitian
mass matrices of the multiplets,
which can be written as
\begin{equation}\label{7}
{\mathds{M}}^{(i)}=
\sum\limits_{l=1}^{3}{m_{l}^{(i)}}{\mathds{P}}_{l}^{(i)}=
\sum\limits_{l=1}^{3}{m_{l}^{(i)}}\left(n^{(i)}_{l}\otimes \na{i}{l}\right).
\end{equation}
\noindent Here $m_{l}^{(i)}$ are eigenvalues of the mass matrices,
which have the meaning of
masses of the multiplet  components (it is natural to treat $m_{l}^{(i)}$ as positive numbers). The matrices ${\mathds{P}}_{l}^{(i)}$ are orthogonal projectors on the subspaces of states with these masses. These operators
can be expressed via eigenvectors $n^{(i)}_{l}$ of the mass matrices
(the asterisk denotes
complex conjugation). The eigenvectors $n^{(i)}_{l}$  form  orthonormal bases of a three-dimensional vector space over the field of complex numbers and can be obtained from the standard basis
\begin{equation}\label{00001}
e_{1}=\left(
          \begin{array}{c}
            1 \\
            0 \\
            0 \\
         \end{array}
        \right),\quad
e_{2}=\left(
          \begin{array}{c}
            0 \\
            1 \\
            0 \\
          \end{array}
        \right),\quad
e_{3}=\left(
         \begin{array}{c}
            0 \\
            0 \\
            1 \\
          \end{array}
       \right),
       \end{equation}
\noindent with the help of unitary matrices
${\mathds{V}}^{(i)}$: $n^{(i)}_{l}={\mathds{V}}^{(i)}e_{l}$.

This model  provides  mixing of fermion generations. Moreover, in such a model  the phenomenon of particle oscillations (in particular, neutrino oscillations) arises.  The matrix of the
mixing coefficients for  quarks is an
analog to the Cabibbo--Kobayashi--Maskawa (CKM) matrix
\begin{equation}\label{l4}
{\mathds{U}}^{\mathrm{CKM}}={\mathds{V}}^{(u)\dag}
{\mathds{V}}^{(d)},
\end{equation}
\noindent and for  leptons it is  an
analog to the  Pontecorvo--Maki--Nakagawa--Sakata (PMNS) matrix
\begin{equation}\label{l5}
{\mathds{U}}^{\mathrm{PMNS}}={\mathds{V}}^{(l)\dag}{\mathds{V}}^{(\nu)}
.\end{equation}
\noindent Since
the procedure of quantization of this model is well defined,
it is possible to obtain the Dyson decomposition,
which enables one to construct the perturbation theory in the interaction representation.
As a consequence,  we can calculate  the contributions due
to radiative corrections  using the regular diagram technique.

Let us use this model to study  neutrino interaction with matter.
As it was proposed in the paper by Wolfenstein \cite{Wolf},
if  matter density is high enough for considering neutrino interaction
with the background fermions as coherent, it is possible to describe
neutrino interaction with  matter by an effective potential.
The origin of this effective potential is forward
elastic scattering of neutrinos on the fermions of the matter.
In the framework of the scheme under consideration we can get the
explicitly covariant description of the neutrino interaction with
dense matter
based on the analog to the Dirac--Schwinger equation of quantum electrodynamics
(see, e.g., \cite{BSh}).

For greater clarity, consider a neutrino propagation in the environment consisting of electrons, protons and neutrons $(e,p,n)$, assuming that the density of neutrino flux is small. Then in the lowest order of the perturbation theory only two  diagrams shown in figures 1,2  contribute to the mass operator.
\vspace{10pt}
\begin{figure}[H]
\begin{center}
\begin{fmffile}{chm1}
\begin{fmfgraph*}(200,60)
\fmfleft{i1} \fmfright{o1}
\fmf{dbl_plain,label=${{\bb e}}$,l.side=left,left=1,tension=.5}{t1,t2}
\fmf{fermion,label=${\nu}$,l.side=right}{i1,t1}
\fmf{photon,label=${W}$,l.side=right}{t1,t2}
\fmf{fermion,label=${\nu}$,l.side=right}{t2,o1}
\fmfdot{t1,t2}
\end{fmfgraph*}
\end{fmffile}
\caption{Contribution to the mass operator due to the interaction via charged currents.}
\end{center}
\end{figure}
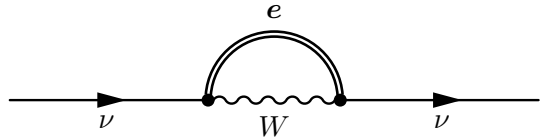
\vspace*{30pt}
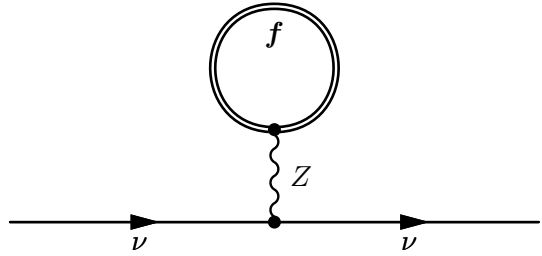
\begin{figure}[H]
\begin{center}
\begin{fmffile}{chm4}
        \begin{fmfgraph*}(200,70)
          \fmfleft{i1} \fmfright{o1}
          \fmf{fermion,label=${\nu}$,l.side=right}{i1,t1}
          \fmf{fermion,label=${\nu}$,l.side=right}{t1,o1}
          \fmffreeze
          \fmftop{i2}
          \fmf{fermion,label=${\nu}$,l.side=right}{i1,t1}
          \fmf{fermion,label=${\nu}$,l.side=right}{t1,o1}
          \fmf{photon,label=${Z}$,l.side=right}{t1,i2}
          \fmf{dbl_plain,label=${{\bb f}}$,l.side=left,left=1,tension=.5}{i2,i2}
        \fmfdot{t1,i2}
        \end{fmfgraph*}
      \end{fmffile}
\vspace*{-10pt}
\caption{Contribution to the mass operator due to the vacuum polarization.}
\end{center}
\end{figure}
\vspace{-10pt}
Therefore, the analog to the Dirac--Schwinger equation for neutrino
takes the form
\begin{multline}\label{ll4}
\left( \ii\gamma^\mu\partial_\mu{\mathds{I}}  -{\mathds{M}}^{(\nu)}\right)
{\varPsi}^{(\nu)}(x)\\
+ \ii \,\frac{g^{2}}{8}\int\!\! d^{4}y \,\gamma^{\mu} (1+\gamma^{5}) S^{(e)}(x,y|{\sf{g}})\gamma^{\nu}(1+\gamma^{5}) D^{W}_{\nu\mu}(y-x)
{\varPsi}^{(\nu)}(y)\\
- \ii \frac{g^{2}}{8\cos^{2}\theta_{\mathrm{W}}}\,{\mathds{I}}
\!\!\sum\limits_{i=e,u,d}\int d^{4}y \,\gamma^{\mu}(1+\gamma^{5}) D^{Z}_{\mu\nu}(x-y)\\ \times\Sp\left\{\gamma^{\nu}
\left(T^{(i)}-2Q^{(i)}\sin^{2}
\theta_{\mathrm{W}}+T^{(i)}\gamma^{5}\right)S^{(i)}(y,y|{\sf{g}})\right\}
{\varPsi}^{(\nu)}(x) = 0.
\end{multline}
\noindent  Here $D^{W}_{\mu\nu}(x-y)$ and  $D^{Z}_{\mu\nu}(x-y)$ are
the causal Green functions of free ${W}$ and ${Z}$ bosons
consequently. And  $S^{(i)}(x,y|{\sf{g}})$ are the causal Green
functions of the fermion multiplets in the
real-time
formalism
taking into account  external conditions $\sf{g}$, i.e. the
temperature and the chemical potential of the background (see, e.g., \cite{levinson}, \cite{zhukovsky}, and the references  cited
therein).

For relatively small neutrino energies when ${\cal E}_{\nu}\ll M_{W}^{2}/{\cal E}_{F}\lesssim M_{W}^{2}/T_{f}$, ${\cal E}_{F} \lesssim T_{f}\ll M_{W}$, where
${\cal E}_{F}, T_{f}$ are the Fermi energy and the temperature of the background fermions (see, e.g., \cite{eminov} and the references  cited therein),
it is possible to use the Fermi approximation. Then
\begin{equation}\label{ll5}
D^{W}_{\mu\nu}(x-y)\approx \frac{g_{\mu\nu}}{M_{W}^{2}} \delta(x-y),\quad D^{Z}_{\mu\nu}(x-y)\approx \frac{g_{\mu\nu}}{M_{Z}^{2}}\delta(x-y),
\end{equation}
\noindent  and Eq. \eqref{ll4} takes the form
\begin{multline}\label{l6}
\left( \ii\gamma^\mu\partial_\mu{\mathds{I}}  -{\mathds{M}}^{(\nu)}\right){\varPsi}^{(\nu)}(x)
+ \ii \frac{G_{\mathrm{F}}}{\sqrt{2}}\Big\{\gamma^{\mu}(1+\gamma^{5}) S^{(e)}(x,x|{\sf{g}})\gamma_{\mu}(1+\gamma^{5})
\\
- \gamma^{\mu}(1+\gamma^{5})\,{\mathds{I}}\!\!\!\sum\limits_{i= e,u,d} \!\!\! \Sp\left\{\gamma_{\mu}\left(T^{(i)}-2Q^{(i)}\sin^{2}
\theta_{\mathrm{W}}+T^{(i)}\gamma^{5}\right)S^{(i)}(x,x|{\sf{g}})\right\}\!\!\Big\}\,
{\varPsi}^{(\nu)}(x) = 0.
\end{multline}

The imaginary parts of the Green functions are, in fact,
the density matrices of the fermions of the external medium, i.e.
$S^{(i)}(x,x|{\sf{g}})\Rightarrow -\ii\varrho^{(i)}(x,x|{\sf{g}}).$
After summation over the quantum numbers of the background fermions,
the density matrices take the form that is well known from general considerations \cite{Michel}.
Assuming that now it is  necessary to consider constituent parts of the medium as the components  of the multiplets, we have
\begin{equation}\label{l7}
\varrho^{(i)}(x,x|{\sf{g}})=\sum\limits_{l=1,2,3}{\mathds
P}^{(i)}_{l}\frac{{\sf n}^{(i)}_{l}}{4p^{0{(i)}}_{l}}
(\gamma_{\alpha}p^{\alpha{(i)}}_{l}+m_{l}^{(i)})
(1-\gamma^{5}
\gamma_{\alpha}s^{\alpha{(i)}}_{l}),
\end{equation}
\noindent where ${\sf n}^{(i)}_{l}$ is
the number density of the multiplet components, and $p^{\alpha{(i)}}_{l}, s^{\alpha{(i)}}_{l}$ are averaged 4-momentum and 4-polarization of the multiplet components respectively.

As a result of elementary calculations we get
\begin{equation}\label{l8}
\displaystyle\gamma^{\mu}(1+\gamma^{5})
\varrho^{(e)}(x,x|{\sf{g}})\gamma_{\mu}(1+\gamma^{5})= - {\mathds
P}^{(e)}\left({j^{\alpha{(e)}}}
-{\lambda^{\alpha{(e)}}}\right)
\gamma_{\alpha}(1+\gamma^{5}),\phantom{ssssssssddddddd}
\end{equation}
\vspace{-25pt}
\begin{multline}\label{l9}
\displaystyle\gamma^{\mu}(1+\gamma^{5})
\Sp \left\{\gamma_{\mu}\left(T^{(i)}-2Q^{(i)}\sin^{2}
\theta_{\mathrm{W}}+T^{(i)}\gamma^{5}\right)
\varrho^{(i)}(x,x|{\sf{g}}))\right\}
 \\ \vspace{35pt}
= \displaystyle\sum\limits_{l=1,2,3}\left({j^{\alpha{(i)}}_{l}}
\left(T^{(i)}-2Q^{(i)}\sin^{2}
\theta_{\mathrm{W}}\right)-
{\lambda^{\alpha{(i)}}_{l}}T^{(i)}\right)\gamma_{\alpha}(1+\gamma^{5}).
\end{multline}
\noindent Here
\begin{equation}\label{l10}
j^{\alpha{(i)}}_{l}={\sf n}^{(i)}_{l}\frac{p^{\alpha{(i)}}_{l}}
{p^{0{(i)}}_{l}}=\{\bar{{\sf n}}^{(i)}_{l} u^{0{(i)}}_{l},\bar{{\sf n}}^{(i)}_{l}{\bf{u}}^{{(i)}}_{l}\}
\end{equation}
\noindent are the  currents, and
\begin{equation}\label{l11}
\lambda^{\alpha{(i)}}_{l} ={\sf n}^{(i)}_{l}\frac{s^{\alpha{(i)}}_{l}}{p^{0{(i)}}_{l}}=
\left\{\bar{{\sf n}}^{(i)}_{l}
({\bb{\zeta}}^{(i)}_{l}{\bf{u}}^{{(i)}}_{l}), \bar{{\sf n}}^{(i)}_{l}\left({\bb{\zeta}}^{(i)}_{l} + \frac{
{\bf{u}}^{{(i)}}_{l} ({\bb{\zeta}}^{(i)}_{l}{\bf{u}}^{{(i)}}_{l})}
{1+u^{0{(i)}}_{l}}\right)\right\}
\end{equation}
\noindent are the polarizations of the background fermions.
In these formulas $\bar{{\sf n}}^{(i)}_{l}\!$ and $\!{\bb{\zeta}}^{(i)}_{l}
(0\leqslant\! |{\bb{\zeta}}^{(i)}_{l} |^2\! \leqslant \!1)$ are
the number density and the mean value of the polarization vector of
the background fermions  in the center-of-mass system of matter, respectively. In this reference frame the mean momentum of the fermions
is equal to zero.
The 4-velocity of this reference frame is denoted  as $u^{\mu}_{l} =\{u^{0{(i)}}_{l},{\bf{u}}^{{(i)}}_{l}\}$.
It is significant that only $j^{\alpha{(i)}}_{l}$ and
$\lambda^{\alpha{(i)}}_{l}$ characterize
medium as a whole.

Let us introduce effective 4-potentials. The potential
\begin{equation}\label{l12}
f^{\alpha{(e)}} =\sqrt{2}{G}_{{\mathrm F}}\left({j^{\alpha{(e)}}}
-\lambda^{\alpha{(e)}}\right)
\end{equation}
\noindent determines the neutrino interaction
with electrons  via the charged currents, while the potential
\begin{equation}\label{l14}
f^{\alpha}_{\mathrm N} =\sqrt{2}{G}_{{\mathrm F}}\sum\limits_{i=e,p,n}
\left({j^{\alpha{(i)}}}
\left(T^{(i)}-2Q^{(i)}\sin^{2}
\theta_{\mathrm{W}}\right)-
{\lambda^{\alpha{(i)}}}T^{(i)}\right)
\end{equation}
\noindent determines the neutrino interaction
with all background fermions via the neutral currents.
Using these potentials we can write the  effective equation,
 which describes neutrino oscillations in matter, in the form
\begin{equation}\label{l15}
\bigg(i\gamma^{\mu}{\partial}_{\mu}{\mathds I}  - {\mathds{M}}^{(\nu)}-
\frac{1}{2}
\gamma^{\alpha}f_{\alpha}^{(e)}(1 + \gamma^5){\mathds P}^{(e)} -\frac{1}{2}
\gamma_{\alpha}f^{\alpha}_{{\mathrm N}}(1 + \gamma^5)\,{\mathds I} \!\bigg)\,{\varPsi}^{(\nu)}(x) = 0.
\end{equation}

This equation generalizes the equation that was used in  papers \cite{RT}, \cite{StT}
for describing a neutrino spin precession in dense matter.
If the effective potentials
are independent of the event space  coordinates, we can write solutions to Eq. \eqref{l15} in the form of matrix exponentials, using the method developed in  papers \cite{SpinLight}, \cite{arlomur}.

The author is grateful to A.V. Borisov, P.A. Eminov, I.P. Volobuev, A.E. Shabad, and V.Ch. Zhukovsky for fruitful discussions.


\begin{thebibliography}{99}



\bibitem{Lobanov_2} A. E. Lobanov, arXiv:1507.01256[hep-ph].



\bibitem{pdg}

K. Olive {\it et al.} (Particle Data Group), \emph {Chin. Phys. C},  {\bf 38}:9 (2014), 090001.



\bibitem{Wolf} L. Wolfenstein, Phys. Rev. D, {\bf 17}:9 (1978), 2369-2374.


\bibitem{BSh} N. N. Bogoliubov  and D. V. Shirkov, {\it Introduction to Theory of Quantized Fields}, John Wiley, New York, 1979.

\bibitem{levinson} E. J. Levinson, Phys. Rev. D, {\bf 31}:12 (1985), 3280-3284.

\bibitem{zhukovsky} A. V. Borisov, A. S. Vshivtsev, V. Ch. Zhukovskii, P.A. Eminov,  Usp. Fiz. Nauk, {\bf 167}:3 (1997),  241-267 [Physics-Uspekhi, {\bf 40}:3 (1997), 229-255].


\bibitem{eminov} P. A. Eminov, Zh. Eksp. Teor. Fiz, {\bf 149}:1 (2016),  76-92 [JETP, {\bf 122}:1 (2016), 63-77].



\bibitem{Michel} L. Michel and A. S. Wightman, Phys. Rev., {\bf 98}:4 (1955), 1190.







\bibitem{RT}
 A. E. Lobanov, Dokl. Akad. Nauk Ser. Fiz. {\bf 402}:4 (2005), 475-478  [Dokl. Phys. {\bf 50}:6 (2005), 286-289] (arXiv:hep-ph/0411342).



\bibitem{StT} A. Studenikin and A. Ternov, Phys. Lett. B {\bf 608}:1-2 (2005), 107-114  (arXiv:hep-ph/0412408).



\bibitem{SpinLight} A. E. Lobanov, Phys. Lett. B, {\bf 619}:1-2 (2005),
136-144  (arXiv:hep-ph/0506007).



\bibitem{arlomur} E. V. Arbuzova, A. E. Lobanov, and E. M. Murchikova, Phys. Rev. D, {\bf 81}:4 (2010), 045001,
16 pp. (arXiv:0903.3358[hep-ph]).

\end{thebibliography}
\end{document}